\documentstyle[12pt]{article}
\date{}
\def\be{\begin{equation}}
\def\ee{\end{equation}}
\def\ba{\begin{eqnarray}}
\def\ea{\end{eqnarray}}
\title{Is the Deformation Parameter in $q$-Rotor Model \\ 
Really Phenomenological ?}
\author{{\it Ramandeep S. Johal}, \\
Department of Physics, \\ Panjab University, Chandigarh - 160014,\\ India.}
\begin{document}
\maketitle
\vskip 18pt
\textwidth 15cm
\begin{abstract}
We cast the $q$-rotor in the framework of Barnett-Pegg theory for
rotation angle, whose underlying algebra is $SU_q(2)$. A new method to 
fix the deformation parameter
from the theory is suggested. We test our ideas by fitting rotational
spectra in deformed even-even and superdeformed nuclei. The results are
in good agreement with the previous phenomenological applications of
$q$-rotor model.
\end{abstract}
\newpage
In recent years, quantum algebras have been extensively applied in
nuclear physics ( for a review, see ref. [1]). In particular, quantum
algebra $SU_q(2)$ has been used in describing the rotational spectra of
deformed even-even nuclei [2] and superdeformed nuclei [3]. The basic
model adopted in these studies is $q$-rotor, whose hamiltonian is
written in terms of second order casimir of $SU_q(2)$ . Using this
model, improved fits are obtained for the rotational spectra. But the
approach has been mainly phenomenological, where $SU_q(2)$  dynamical
symmetry of the system was rather assumed. In any case, some physical
significance could always be ascribed to the deformation parameter.
 For example, $q$ has been related to the softness parameter of
variable moment of inertia model [4].

On a different side, a few years back, Pegg and Barnett proposed a
solution to the long standing phase operator problem of quantum harmonic
oscillator [5]. A similar formalism also exists for the rotation angle
operator [6]. However, although much work has been done related to
phase operartor for electromagnetic fields [7,8], very little has been
discussed regarding the rotation angle described in this formalism. A
major reason for this indifference appears to be that `angular velocity'
was ill-defined in this framework [9]. Recently, two new solutions have
been suggested to remove this malady [10,11]. So in our opinion,
Barnett-Pegg (BP) formalism is a valid approach to
the theory of rotation angle in quantum mechanics.

In this letter, we seek a theoretical basis for $q$-rotor model.
The underlying framework for our theory is BP formalism.
Very briefly, in this formalism, a complete set of orthogonal angle
states is defined in $(2l+1)$-Hilbert,
\be
|\theta_n\rangle=\frac{1}{\sqrt{2l+1}}\sum_{m=-l}^{l} exp(-im\theta_n)
|m\rangle
\ee
where $\theta_n=\theta_0 +\frac{2\pi n}{2l+1}\; (n= 0,1,..,2l)$. $m$ is the
eigenvalue of $J_z$, component of angular momentum along $z$-axis. 
The  unitary operator
\be
exp(\pm i\Phi) = exp \bigl(\pm i
\sum_{n=0}^{2l}\theta_n|\theta_n\rangle\langle\theta_n|\bigr)
\ee
corresponds to the hermitian angle operator, $\Phi$. It can be
said that the above formalism has inherent $q$-deformed structure [10],
if we identify $q=exp(-i\frac{2\pi}{2l+1})$. Then the operators
$q^{J_z/\hbar}$ and $e^{i\Phi}$ satisfy
\be
q^{J_z/\hbar}e^{i\Phi}=q \;  e^{i\Phi} q^{J_z/\hbar}
\ee
which is the quantum plane condition in the discrete angular
momentum-angle phase space [12]. These operators show cyclic displacement
property
\be
e^{i\Phi}|m\rangle =|m+1\rangle
\ee
\be
e^{i\Phi} |l\rangle = e^{i(2l+1)\theta_0} |-l\rangle.
\ee
$e^{-i\Phi}$ is the lowering operator in this representation. Operator
 $q^{J_z/\hbar}$  has analogus action on $|\theta_n\rangle$ states.
All this implies a periodic lattice structure, the lattice
constant being related to the deformation parameter. It has
been suggested  [13] quite generally, that whenever there is an underlying
lattice or presence of discrete lengths in the system, the related
algebraic structure is a $q$-deformed one.

Thus we note that the algebra of operators in BP-formalism is
$q$-deformed angular momentum algebra or briefly, $SU_q(2)$, with $q$ as
root of unity. For the
purpose of $q$-rotor, it is essential to study the full set of
generators of $SU_q(2)$. The algebra  is given by the following commutation
relations
\ba
~[J_+,J_-] & = & [2J_z]=\frac{q^{2J_z}-q^{-2J_z}}{q-q^{-1}},\\
~[J_z,J_{\pm}] & = & \pm J_{\pm}
\ea
The operators act on the subspace spanned by angular momentum
eigenstates $|jm\rangle$, where the representation label $j$ is very less
than $l$, which denotes the full space. Now for each value of $j$,
operators $J_{\pm}$ act only on $|jm\rangle \;(m=-j, -j+1,....,j)$. 
They do not
mix eigenstates with different values of $j$. Thus each complete set of
$|jm\rangle$ states, serves as a basis for a $(2j+1)$-irrep for $SU_q(2)$. The
action of $J_{\pm}$ is given by
\ba
~J_{\pm}|jm\rangle & = &\hbar \{[j\mp m][j \pm m + 1]\}^{1/2} |jm\pm
1\rangle \\
~J_z|jm\rangle  & = & m\hbar|jm\rangle.
\ea
The second order casimir of this algebra is
\be
C_2(SU_q(2) ) = [J_z][J_z+1] + J_-J_+,
\ee
with eigenvalues
\be
C_2(SU_q(2) )|jm\rangle =[j][j+1]|jm\rangle.
\ee
Now consider a set of states, each labelled by a different value of $j$,
ranging from $j_{min}$ to $j_{max}$, and incresing in equal steps. This
is our {\it system}.  Question is, how much large the total space with
dimension $(2l+1)$ be ? BP theory requires [6] that for physical states
it is finite but may be arbitrarily large, such that $l\gg j$.
We try to set the
lower bound for the $(2l+1)$ value. A natural way is to demand that the
total hilbert space be
sufficiently large to contain all the $(2j+1)$ irreps of the
{\it system}. More precisely, we set
\be
2l+1 = \sum_{j_{min}}^{j_{max}} (2j+1).  \label{size}
\ee
If the above sum is even, then $(2l+1)$ value is the next odd number.
Remember that  we are guessing the minimum size of the total space and
in this way, maximally fix the deformation parameter from the theory.
In short we suggest that {\it the size  (scale) of the embedding space is
determined by the sum total of dimensions of the contained irreps}. (See
also the remark after eq. (\ref{chi}).

Now we turn to the $q$-rotor model. The hamiltonian is given by
\be
H_q = \frac{1}{2I} C_2(SU_q(2) ),
\ee
where $I$ is the moment of inertia. The energy of a state labelled by
quantum number $j$ is then
\ba
E_j  &= & \frac{1}{2I} [j][j+1]  \nonumber \\
     & =  &  \frac{1}{2I} \frac{sin\;(\tau j) sin\;(\tau (j+1))
        }{sin^2\;(\tau)} \label{ener}
\ea
where $\tau =\frac{2\pi}{2l+1}$ and $(2l+1)$ is determined from eq. 
(\ref{size}).

We apply the above ideas to rotational spectra in deformed even-even
nuclei and superdeformed nuclei. We have fitted the parameter $A=1/2I$
in eq. (\ref{ener}) for
some nuclei and the results are tabulated in table 1. The improvement of
fits over the classical $j(j+1)$ formula is indicated by the decrease in
the quantity
\be
\chi ^2 = \sum_{j_{min}}^{j_{max}} \{exp(j)-theo(j)\}^2.\label{chi}
\ee
Before discussing the results, a remark is in place. In the standard BP
aprroach, after the physical quantities have been calculated, $l$ is
made to go to infinity. This is necessary, because only in the
semiclassical limit of large angular momentum, the phase operator
becomes continuous. However, we argue that $q$-rotor is
discussed in angular momentum representaion and we make no attempt to
determine physical quantities in terms of $\Phi$. Thus the angle
eigenvalues remain indeterminate.  Secondly, the square bracket $[j]$, which
generally can also take negative values for $q$ root of unity, always remains
positive here. This is because of $l\gg j$.

It is observed that in general, there is significant decrease in root
mean square deviation showing good improvement of
results over the ordinary rotor (for $^{238}$U, the classical rms
deviation is about 250). The most interesting observation is
that for deformed even-even nuclei, $\tau$ assumes values around 0.03.
This was also found in [2], where $\tau$ was treated as fitting
parameter. The fitted parameter $A=1/2I$ also assumes similar
close values.

In the case of superdeformed nuclei, $\tau$  assumes still smaller
values. In 130 mass region,  $\tau$ $\approx 0.01$. For 150 mass region, 
$\tau$
$\approx 0.004$. Both these results match with the phenomenological
findings of [3]. In 190 mass region, $\tau$  is in between the above two
values. Thus on the average, there is a monotonic decrease of $\tau$  in 
different mass
regions with increase of superdeformation.
Also it has been found on phenomenological basis [2,3], that for
one deformation parameter fitting of $q$-rotor, only $q=e^{i\tau}$ and
not $q=e^{\tau}$ can give improved
fitting of the data. $q$-parameter inbuilt in the BP approach
is already pure phase (modulus unity) and so it is suitable for
describing the $q$-rotor.

Concluding, we have studied $q$-rotor within Barnett-Pegg approach. In
the present work,  a method to fix the deformation parameter {\it a priori}
from the theory has been suggested.
We have applied the idea
to describe rotational spectra in deformed even-even and superdeformed
nuclei. The results are quite close to those of previous phenomenological
approaches to the $q$-rotor problem. Attempt is underway to understand 
in a similar
vein, the rotational molecular spectra.
One may look forward to describe excited collective $\beta-$ and $\gamma-$
bands of even deformed nuclei starting from the above theory. However, as
pointed out in [15], alone $SU_q(2)$ symmetry may not be sufficient as 
these excited bands
involve non-rotational degrees of freedom also.
It may be possible to formulate phase theories more general than BP theory,
which would incorporate deformed symmetries other than $SU_q(2)$.
\newpage
\section*{References}
\begin{enumerate}
\item{} D. Bonatsos, C. Daskaloyannis, P. Kolokotronis and D. Lenis,
 Quantum Algebras in Nuclear Structure, nucl-th/9512017.
\item{} P.P Raychev, R.P. Roussev and Yu F. Smirnov, J. Phys. G: Nucl.
Part. Phys., {\bf 16} (1991) L137.
\item{} D. Bonatsos, S.B. Drenska, P.P. Raychev, R.P. Roussev and Yu F.
Smirnov, J. Phys. G: Nucl.
Part. Phys., {\bf 17} (1991) L67.
\item{} D. Bonatsos, E.N. Argyres, S.B. Drenska, P.P. Raychev, R.P. Roussev 
and Yu F.  Smirnov, Phys. Lett. B, {\bf 251} (1990) 477.
\item{} D.T. Pegg and S.M. Barnett, Europhys. Lett. {\bf 6} (1988) 483;
Phys. Rev. A {\bf 39} (1989) 1665.
\item{} S.M. Barnett and D.T. Pegg, Phys. Rev. A, {\bf 41} (1990) 3427.
\item{} A. Luks and V. Perinova, Physica Scripta: Special Issue devoted
to: Quantum Phase and Phase Dependent Measurements, {\bf T48} (1993) 94.
\item{} R. Lynch, Phys. Rep. {\bf 256} (1995) 367.
\item{} D. Loss and K. Mullen, J. Phys. A: Math. and Gen., {\bf 25} (1992) 
L235.
\item{} S. Abe, Phys. Rev. A, {\bf 54} (1996) 93.
\item{} R.S. Johal, Angular Velocity Operator in Barnett-Pegg Formalism,
quant-ph/9707029, submitted to Phys. Rev. A.
\item{} A. Vourdas, Phys. Rev. A, {\bf 41} (1990) 1653; {\bf 43} (1991)
1564.
\item{} E. Celeghini, S. De Martino, S. De Siena, M. Rasetti and G.
Vitiello, Annal. Phys. {\bf 241} (1995) 50.
\item{} B. Firestone and B. Singh, Table of Superdeformed Nuclear Bands
and Fission Isomers, June 1994, LBL-35916.
\item{} N. Minkov, S.B. Drenska, P.P. Raychev, R.P. Roussev and D. Bonatsos,
J. Phys. G: Nucl. Part. Phys., {\bf 22} (1996) 1633.
\end{enumerate}
\newpage
Table~1.~~~Fitting of eq. (14) with the theoretically determined value of 
deformation
parameter $\tau$. Upper box is for deformed even-even nuclei while lower box
shows superdeformed nuclei. Data for deformed nuclei
is as used in [3,4]. For superdeformed nuclei, data is taken from ref.
[14]. $A$ is in keV and $\chi^2$ is measured in $(MeV)^2$.
\vskip 18pt
\hskip 1.5cm
\begin{tabular}{|l|c|c|c|}
\hline

~~~Nucleus~~~ &~~~  $\tau$~~~ &~~~  $A=1/2I$~~~    &~~~   $10^3\;\chi^2$~~~ \\
\hline
$^{162}$Dy   & .0332 & 12.81  &  2.22 \\
$^{174}$Yb   & .0273 & 12.26  &  2.81 \\
$^{176}$Yb   & .0332 & 13.32  &  2.98 \\
$^{178}$Hf   &  .0332  &  14.05  &  37.66  \\
$^{232}$U    &  .0273  &  7.15  &   8.19  \\
$^{238}$U    &  .0145  &  6.16  &  162.2  \\
$^{238}$Pu   &  .0332  &  7.39  &  0.5  \\
\hline\hline
$^{130}$La   & .0123 & 11.43 & 4.07 \\
$^{134}$Nd   & .0103 & 9.80 & 1.83 \\
$^{136}$Nd   &  .0095  & 10.10 &  2.0 \\
$^{146}$Gd-1 & .0041 & 6.33 & 5.93 \\
$^{146}$Gd-2 &  .0045 &  6.48 &  19.10  \\
$^{150}$Tb   & .0039  &  6.26  & 13.77 \\
$^{152}$Dy-1   & .0038  &  6.27  &  10.02  \\
$^{190}$Hg     &  .0074  &  5.43  &  5.86 \\
$^{192}$Hg   &  .0057  &  5.10  &  8.87  \\
$^{194}$Hg-1 &.0058   &   5.05  &   8.17  \\
$^{194}$Hg-2 & .0074  &  5.09  &   0.98   \\
$^{194}$Hg-3 &  .0075  &  5.11  &  0.99  \\
\hline
\end{tabular}
\end{document}